\documentclass[12pt]{iopart}
\eqnobysec
\usepackage{iopams} 
\usepackage{epsfig}
\usepackage{amssymb}
\begin{document}

\title[Measures of non-Gaussianity for one-mode field states ]{Measures of non-Gaussianity for one-mode field states}

\author{Iulia Ghiu{$^1$},
Paulina Marian{$^{1,2}$} 
and Tudor A. Marian{$^1$}}

\address{$^1$Centre for Advanced  Quantum Physics,
Department of Physics, University of Bucharest, 
R-077125 Bucharest-M\u{a}gurele, Romania}

\address{$^2$ Department of Physical Chemistry,
University of Bucharest, Boulevard Regina Elisabeta 4-12, R-030018  Bucharest, Romania}
\ead{paulina.marian@g.unibuc.ro}
\begin{abstract}We introduce and investigate a distance-type 
measure of non-Gaussianity based on the quantum fidelity. 
This new measure can readily be evaluated 
for all pure states and mixed states that are diagonal in the Fock basis. 
In particular, for an $M$-photon added thermal state, an analysis of the Bures degree of non-Gaussianity is made in comparison with two previous measures built with the Hilbert-Schmidt metric and the relative entropy. We obtain a compact analytic formula 
for the Hilbert-Schmidt non-Gaussianity measure and find a good consistency of the three examined measures.
\end{abstract}


\maketitle

\section{Introduction}
   Originally, non-Gaussian states
were studied in quantum optics due to some non-classical properties: photon antibunching, quadrature or amplitude-squared
squeezing, oscillations of the photon-number distribution. More generally, non-classicality was defined as the non-existence 
of the $P$ representation as a well-behaved function. A review
of the efforts made on these lines of research can be found 
in Ref.\cite{1}. In the pure state case there exists a connection between non-classicality and the non-Gaussian charater 
of the density operator. Indeed, Cahill \cite{Cah}  proved that the only 
pure states that are classical are the coherent ones: all other classical states are mixtures. Therefore, all pure non-Gaussian states are non-classical. On the other hand, according to Hudson's theorem \cite{Hud},  all pure states
with negative Wigner function are non-Gaussian. 
Signatures of non-classicality
could be thus identified  through the negativity of the Wigner
function. It was shown that this holds for mixed non-Gaussian states as well \cite{KZ}.
Interest in the non-Gaussian states  has recently emerged
in quantum information processing.
It was realized that non-Gaussian resources and operations could be more performant in some quantum  protocols such as teleportation \cite{Opat,Paris1,Il} and cloning \cite{Cerf1}. To understand to what extent non-Gaussianity could be a resource in such cases, some distance-type measures of this property were proposed \cite{P2,P3} following one of the patterns
\begin{equation}
\delta[\hat \rho] \sim {d}^2(\hat \rho, {\hat \rho}_G) \qquad {\rm or} \qquad \delta[\hat \rho]:={d}(\hat \rho, {\hat \rho}_G), \label{1}
\end{equation}
where ${\hat \rho}_G$ is the Gaussian state having the same average displacement and covariance matrix as the given
state $\hat \rho$. 
In Refs.\cite{P2,P3,P33}, Genoni
{\em et al.} employed as distance $d$ the 
Hilbert-Schmidt metric and the relative entropy: they defined the 
Hilbert-Schmidt measure of non-Gaussianity \cite{P2},
\begin{equation}
\delta_{HS}[\hat \rho]:=\frac{{d}^2_{HS}(\hat \rho, {\hat \rho}_G)}
{2\,{\rm Tr} (\hat \rho^2)}\, , \label{hs}
\end{equation}
as well as the relative entropy of non-Gaussianity \cite{P3},
\begin{equation}\delta_{RE}[\hat \rho]:={\cal S}(\hat \rho|
{\hat \rho}_G):={\rm Tr}(\hat \rho\ln {\hat \rho})
-{\rm Tr}(\hat \rho\ln {\hat \rho}_G). \label{re}
\end{equation}
 Interestingly, non-Gaussianity in terms of relative entropy \cite{P3} was experimentally measured for single-photon added coherent states \cite{P4}.
Another approach to non-Gaussianity was based on $Q$ function  and lead to a measure expressed by the difference between the classical (Wehrl) entropies of the Gaussian state ${\hat \rho}_G$ and the given non-Gaussian state $\hat \rho$ \cite{Simon}.

In this paper we introduce a measure of non-Gaussianity 
of a single-mode state $\hat \rho$ in terms of its Bures distance to the Gaussian state ${\hat \rho}_G$ having the same first- 
and second-order moments of the canonical quadrature operators. 
In other words, our definition is of the type \ (\ref{1}) and uses a well-known metric related to the  fidelity between two quantum states \cite{Uhl}.  Fidelity-based metrics have proven to be fruitful in quantum optics and quantum information as measures of nonclassicality \cite{PTH02},  entanglement \cite{PTH,PTHa,PTHb}, 
 and polarization \cite{Bj,Bja,lu,OC,IGPT}. We also intend to compare the three above-mentioned distance-type measures in analyzing 
the non-Gaussianity for a definite class of one-mode states.

The plan of our paper is as follows. In Sec. 2 we  introduce and examine the Bures degree of non-Gaussianity. We insist on its advantageous form for pure states and for mixed ones 
that are diagonal in the Fock basis. Section 3 investigates an interesting mixed state of this type which is important 
for experiments: a thermal state with $M$ added photons.  
We here give a compact analytic form of $\delta_{HS}$, while  
$\delta_{RE}$ and $\delta_{F}$ are expressed in terms of two series
which have to be summed numerically. Section 4 is devoted to 
a discussion of our numerical results, making a comparison 
between the three above-mentioned non-Gaussianity measures and analyzing their consistency.

\section{Bures measure of non-Gaussianity}
 Following the definition\ (\ref{1}) one can take advantage of the
distinguishability properties possessed by the distance ${d}$  in order to get reliable values for non-Gaussianity. For further convenience we write down the previously defined degrees \cite{P2,P3}. On the one hand, Eq.\ (\ref{hs}) gives an easily computable 
expression:
\begin{equation} 
\delta_{HS}[\hat \rho]=\frac{{\rm Tr } [(\hat \rho-\hat \rho_G)^2]}{2 {\rm Tr } (\hat \rho^2)}=\frac{1}{2}\left[1+\frac{{\rm Tr } (\hat \rho_G^2)-2 {\rm Tr} (\hat \rho_G \hat \rho)}{{\rm Tr } (\hat \rho^2)}\right]. \label{hs1}
\end{equation}
On the other hand, despite its not being a true distance, 
the relative entropy is acceptable and used as a measure of distinguishability between two quantum states. Moreover, recall that 
the relative entropy of non-Gaussianity, Eq.\ (\ref{re}), 
reduces to 
\begin{equation} 
\delta_{RE}[\hat \rho]={\cal S}(\hat \rho_G)-{\cal S}(\hat \rho),
\label{re1}
\end{equation}
where ${\cal S}(\hat \rho):=-{\rm Tr}(\hat \rho\ln {\hat \rho})$ 
is the von Neumann entropy of the state $\hat \rho$.
Invariance properties of the degrees \ (\ref{hs1}) and \ (\ref{re1})
were discussed in detail in Refs.\cite{P2,P3,P33}.

We now define a fidelity-based degree of non-Gaussianity 
\begin{equation} 
\delta_{F}[\hat \rho]:=\frac{1}{2}\;{d}^2_{B}(\hat \rho, {\hat \rho}_G)=1-\sqrt{{\cal F}(\hat \rho,\hat \rho_G )}. \label{bu}
\end{equation}
The explicit expression of the fidelity between the states 
$\hat \rho_1$ and $\hat \rho_2$ was written by Uhlmann \cite{Uhl}:
\begin{equation} 
{\cal F}(\hat \rho_1, \hat \rho_2 )=\left\{{\rm Tr}
[(\sqrt{\hat \rho_1}\hat \rho_2 \sqrt{\hat \rho_1})^{1/2}]\right\}^2.\label{F}
\end{equation}
As seen in Eq.\ (\ref{bu}), fidelity is tightly related to the Bures
metric ${d}_{B}$  introduced in Ref.\cite{Bu} on mathematical grounds. Several general properties of our definition\ (\ref{bu}) are listed below as arising from well-known beneficial features of the fidelity \cite{BZ}.
\begin{enumerate}
\item  
The property of the fidelity to vary between $0$ and $1$ implies:
\begin{equation}
\delta_{F}[\hat \rho]=0, \quad {\rm iff} \;\; \hat \rho\;\; 
{\rm is}\;\; {\rm Gaussian}, \label{p1}
\end{equation}

\begin{equation}
0<\delta_{F}[\hat \rho] \leq 1,\quad {\rm otherwise}. \label{p11}
\end{equation}
\item 
If at least one of the states is pure, Eq.\ (\ref{F}) reduces to the usual transition probability ${\rm Tr}(\hat \rho_1 \,\hat \rho_2)$.
Correspondingly, the Bures degree of non-Gaussianity\ (\ref{bu}) 
of a pure state $|\Psi \rangle \langle \Psi|$ is
\begin{equation}
\delta_{F}[|\Psi\rangle\langle \Psi|]= 1-\sqrt{\langle \Psi|\hat \rho_G|\Psi \rangle}. \label{pure}
\end{equation}
\item 
As shown in Refs.\cite{P2,P3,P33}, when $\hat U$ are the unitary operators of the metaplectic representation on the Hilbert space 
of states, then
$ \hat \rho^{\prime}=\hat U \hat \rho \hat U^{\dag} \Longrightarrow (\hat \rho^{\prime})_G=\hat U \hat \rho_G \hat U^{\dag}$ 
and, therefore, according to the invariance of the fidelity under unitary transformations we obtain the identity
\begin{equation}
\delta_{F}[\hat U \hat \rho \hat U^{\dag}]=\delta_{F}[\hat \rho].
\label{p2}
\end{equation}
It follows that $\delta_{F}[\hat \rho]$ does not depend on 
one-mode squeezing and displacement operations. 
\item 
The multiplicativity property of the fidelity  has an interesting consequence on our definition\ (\ref{bu}) for a two-mode product 
state $\hat \rho_1\otimes \hat \rho_2$. Indeed, if  $\hat \rho_2$ 
is a Gaussian state, we get 
${\cal F}(\hat \rho_1 \otimes \hat \rho_2, (\hat \rho_1)_G \otimes \hat \rho_2)=
{\cal F}(\hat \rho_1, (\hat \rho_1)_G)$ and therefore
\begin{equation}
\delta_{F}[\hat \rho_1 \otimes \hat \rho_2]=\delta_{F}[\hat \rho_1].\label{p3}
\end{equation}
\item 
For commuting density operators, $[\hat \rho_1,\hat \rho_2]=\hat 0$,
Eq.\ (\ref{F}) simplifies to 
\begin{equation}{\cal F}(\hat \rho_1, \hat \rho_2)=[{\rm Tr}({\hat \rho_1}^{1/2}
{\hat \rho_2}^{1/2})]^2. \label{p4}
\end{equation}
\end{enumerate}
Let us now remark that the properties\ (\ref{p1}) and\ (\ref{p11}) justify the interpretation of $\delta_{F}[\hat \rho] $ as a degree of non-Gaussianity. At the same time,  properties \ (\ref{p1}),\ (\ref{p2}), and\ (\ref{p3}) of $\delta_{F}[\hat \rho] $ are shared by the non-Gaussianity measures\ (\ref{hs1}) and\ (\ref{re1}) as well \cite{P33}. Note that we do not discuss here the evolution of the
non-Gaussianity of a state under a completely positive map which is expected to be a monotonic one in the cases of the relative-entropy- and fidelity-based degrees \cite{BZ}.

How well these measures discriminate between quantum states in order to be declared good measures of non-Gaussiannity is a complicated question which was already invoked when discussing distance-type measures of non-classicality \cite{PTH02} or entanglement \cite{PTH}.  It is desirable that, for a specific family of non-Gaussian states, any distance-type
degree has a monotonic behaviour with respect to the continuous parameters defining the set of states.  In the case of one-mode states,
we adopt as a
reasonable criterion to verify the appropriateness of the non-Gaussianity measures \ (\ref{hs1}),\ (\ref{re1}), and\ (\ref{bu}), their monotonic behaviour with respect to the average photon number 
of the state $\langle \hat N\rangle$. Another property that one could expect for the three measures is their consistency, namely, their quality to induce the same ordering of non-Gaussianity when considering a specific set of states. It was already shown that relative entropy and Hilbert-Schmidt measures display different ordering for Schr\"odinger cat-like states \cite{P33}. However, conclusions on such important aspects of distance-type degrees 
of non-Gaussianity cannot be drawn in general, but only for special sets of states. This happens because  obtaining compact analytic results is a task that requires diagonalization of the density operator $\hat \rho$ followed by the exact summation of the corresponding power series. Evaluation of $\delta_{HS}[\hat \rho]$  seems to be easier than that of both  $\delta_{RE}[\hat \rho]$ 
and $\delta_{F}[\hat \rho]$. As a matter of fact,
Uhlmann's expression\ (\ref{F}) is not easy to calculate even on finite-dimensional Hilbert spaces. We refer the reader to a recent paper of two of us \cite{PT2012} where the state of the art in evaluating fidelity in the continuous-variable settings is presented.
However, there are some important sets of states for which we can get explicit and relevant results.
In the following we concentrate on such two computable cases.

First, for pure states, Eq.\ (\ref{pure}) shows that $\delta_{F}$ is  state-dependent. This is equally true for $\delta_{HS}$:
\begin{equation}\delta_{HS}[|\Psi\rangle\langle \Psi|]
=\frac{1}{2}\left[1+\Tr (\hat \rho_G^2)
-2 \langle\Psi|\hat \rho_G|\Psi\rangle\right]. \label{hsp}
\end{equation}
Indeed, in order to evaluate $\delta_{F}$ and $\delta_{HS}$, we need 
to determine the reference Gaussian density operator $\hat \rho_G$ and its expectation value in the pure state $|\Psi \rangle\langle \Psi|$. By contrast, the entropic non-Gaussianity measure\ (\ref{re1}) 
of any pure state is a unique function of a single variable, namely, the determinant of the $2\times 2$ covariance matrix of the state, 
$\Delta:=\det (\cal V)$: 
\begin{eqnarray}
\delta_{RE}[|\Psi \rangle\langle \Psi|]=\left( \sqrt{\Delta}+
\frac{1}{2} \right)\ln{\left (\sqrt{\Delta}+\frac{1}{2}\right)}                                   
-\left( \sqrt{\Delta}-
\frac{1}{2} \right)\ln{\left (\sqrt{\Delta}-\frac{1}{2} \right)}. \nonumber\\
\label{rep}
\end{eqnarray}
It is worth mentioning, however, that the Fock states are the only 
pure states for which both the Bures and Hilbert-Schmidt degrees 
of non-Gaussianity\ (\ref{pure}) and\ (\ref{hsp}) depend solely
on the parameter $\Delta$. Let us consider a number state 
$|M \rangle\langle M|$. The associated Gaussian state ${\hat \rho}_G$ is a thermal one with the mean occupancy $\langle \hat N \rangle=M$.
Equations\ (\ref{pure}) and\ (\ref{hsp}) give, respectively,
the formulae: 
\begin{equation}
\delta_{F}[|M\rangle\langle M|]=1-\sqrt{\frac{M^{M}}
{(M+1)^{M+1}}}, \label{MB}
\end{equation}
and
\begin{equation}
\delta_{HS}[|M\rangle\langle M|]=\frac{M+1}{2M+1}-\frac{M^{M}}
{(M+1)^{M+1}}, \label{MHS}
\end{equation}
with $\sqrt{\Delta}=M+\frac{1}{2}$. Note that Eq.\ (\ref{MHS}) 
was already derived in Ref.\cite{P2}. Owing to the invariance 
property\ (\ref{p2}), the expression\ (\ref{MB}) coincides with
the Bures degree of non-Gaussianity of squeezed or/and displaced number states. 

Second, for any mixed Fock-diagonal state,
\begin{equation}
\hat \rho=\sum_{l=0}^{\infty} p_l \, 
|l\rangle \langle l| \quad {\rm with} \quad
\sum_{l=0}^{\infty}  p_l=1, \label{diag}
\end{equation}
the Gaussian reference state $\hat \rho_G$ is a thermal state 
with the same mean occupancy $\langle \hat N \rangle=\sum_l l\,p_l$. We denote $\sigma:=\langle \hat N \rangle/(\langle \hat N \rangle+1)$ and write its spectral expansion:
\begin{equation}
\hat \rho_G=\sum_{l=0}^{\infty} s_l \,
|l\rangle \langle l|  \quad {\rm with}  \quad
s_l =\frac{1}{\langle \hat N \rangle+1}\, \sigma^l. \label{diag1}
\end{equation}
The corresponding Hilbert-Schmidt and entropic non-Gaussianity
measures were written in Refs.\cite{P2,P3}. For further use, 
we cast the Hilbert-Schmidt measure\ (\ref{hs1}) into a slightly modified form:
\begin{eqnarray}
\delta_{HS}[\hat \rho]&=&\frac{1}{2}\left[1+\frac{\sum_{l}(s_l^2-2 s_l\,p_l) }{\sum_{l}p_l^2}\right]\nonumber\\  &&=\frac{1}{2}\left[1+\frac{1}{\sum_{l}p_l^2}\left(\frac{1}{2\langle \hat N 
\rangle+1}-\frac{2}{\langle \hat N \rangle+1}
{\cal G}_{\hat \rho}(\sigma) \right) \right]. \label{hs2}
\end{eqnarray}
Here we have used the purity of the thermal state $\hat \rho_G$ arising from Eq.\ (\ref{diag1}), while  
${\cal G}_{\hat \rho}(y):=\sum_{l}  p_l\, y^l$ is the generating function of the photon-number distribution of the given state 
$\hat \rho$. The relative entropy of non-Gaussianity, 
Eq.\ (\ref{re}), simplifies to:
\begin{eqnarray}
\delta_{RE}[\hat \rho]&=&\sum_{l=0}^{\infty}(p_l\, \ln{p_l}
-s_l\,\ln{s_l})\nonumber\\  && =\sum_{l=0}^{\infty}p_l\ln{p_l}
+(\langle \hat N \rangle+1)\ln (\langle \hat N \rangle+1)
-\langle \hat N \rangle\ln  (\langle \hat N \rangle). \label{re2}
\end{eqnarray}
In the last line we have used the von Neumann entropy of a thermal state.  In this special case we notice the commutation relation 
$[\hat \rho,\hat \rho_G]=\hat 0$,
which allows us the use of Eq.\ (\ref{p4}) to get
\begin{equation}
\delta_{F}[\hat \rho]=1-\sum_{l=0}^{\infty}\sqrt{ p_l\, s_l}.
\label{diag2}
\end{equation}
Note that some important mixed non-Gaussian states have the structure\ (\ref{diag}): phase-averaged coherent states and various
excitations on a thermal state of the type $\hat \rho \sim (\hat a^{\dag})^k \hat a^l \hat \rho_{th}
(\hat a^{\dag})^l \hat a^k $. Here $\hat a$ and $\hat a^{\dag}$ are the amplitude operators of the field mode.

\section{An example: Photon-added thermal states}
In general, the states with added photons are non-classical and non-Gaussian. We choose here to analyze an $M$-photon-added thermal state \cite{AT,JL} as an interesting example of a Fock-diagonal state whose non-classicality was recently investigated in ingenious
experiments \cite{bel,bel1,Kis}. Its density operator is
\begin{equation}
\hat \rho^{(M)}=\frac{1}{M!\, (\bar n+1)^M}\, (\hat a^\dagger )^M\, \hat \rho_{th}\, \hat a^M. \label{PATS}
\end{equation}
Here $M$ is the number of added photons and $\hat \rho_{th}$ is a thermal state whose mean number of photons is denoted by $\bar n$:
\begin{equation}
\hat \rho_{th}=(1-x)\sum_{l=0}^{\infty}\, x^l \, |l\rangle \langle l| \quad {\rm with} \quad
x:=\frac{\bar n}{\bar{n}+1}. \label{diag3}
\end{equation}
Accordingly, the density operator $\hat \rho^{(M)}$, 
Eq.\ (\ref{PATS}), has the following eigenvalues: 
\begin{eqnarray}
p_l:=(\hat \rho^{(M)})_{ll}=\left(
\begin{array}{c} 
l\\ M 
\end{array}
\right)(1-x)^{M+1}{x}^{l-M}, \qquad (l=0,1,2,3,...). \label{pl}
\end{eqnarray}
 The mean occupancy is simply 
\begin{eqnarray}
\langle \hat N \rangle=\bar n (M+1)+M, \label{mo}
\end{eqnarray}
such that the photon-number probabilities of the associated thermal state read:
\begin{eqnarray}
s_l:=(\hat \rho_G^{(M)})_{ll}=\frac{[\bar n (M+1)+M]^l}
{[(M+1)(\bar n+1)]^{l+1}}\, , \qquad (l=0,1,2,3,...). \label{sl}
\end{eqnarray}
The generating function 
\begin{eqnarray}
{\cal G}_{\hat \rho}(y)
:=\sum_{l=0}^{\infty} p_l\, y^l ={y^M}{(1-x)^{M+1}}
\sum_{l=M}^{\infty} \left(
\begin{array}{c} 
l\\ M 
\end{array}
\right) (x y)^{l-M} \label{gf} 
\end{eqnarray}
has a compact form:
\begin{eqnarray}
{\cal G}_{\hat \rho}(y)={y^M}\left(\frac{1-x}{1-x y}\right)^{M+1}
\label{gf1}.
\end{eqnarray}
Hence, the Hilbert-Schmidt scalar product of the states 
$\hat \rho^{(M)}$ and $\hat \rho_G^{(M)}$ is
\begin{eqnarray}
\Tr \left[\hat \rho^{(M)} \hat \rho_G^{(M)}\right]
=\sum_{l=0}^{\infty}\, p_l\, s_l
=\frac{{\langle \hat N \rangle}^M}{(\langle \hat N 
\rangle+\bar n+1)^{M+1}}. \label{rg}
\end{eqnarray}
We have still to evaluate the purity of the state 
$\hat \rho^{(M)}$: 
\begin{eqnarray}
\Tr \left[(\hat \rho^{(M)})^2\right]=\sum_{l=0}^{\infty}\, {p_l}^2
=(1-x)^{2(M+1)}\sum_{l=M}^{\infty} \left(
\begin{array}{c} 
l\\ M 
\end{array}
\right)^2 x^{2(l-M)}. \label{pl2}
\end{eqnarray}
A change of the summation index in Eq.\ (\ref{pl2}) leads us to 
a closed-form result proportional to a Gauss hypergeometric function, Eq.\ (\ref{a1}):
\begin{eqnarray}
\Tr \left[(\hat \rho^{(M)})^2\right]
=(1-x)^{2(M+1)}{_{2}F_{1}}(M+1, M+1; 1 ; x^2). \label{pl3}
\end{eqnarray}
By applying the linear transformation\ (\ref{a2}), we eventually get
the purity as a function of the ratio $x$, in terms of 
a Legendre polynomial\ (\ref{a3}):
\begin{eqnarray}\Tr \left[(\hat \rho^{(M)})^2\right]
=\left(\frac{1-x}{1+x}\right)^{M+1}
{\cal P}_M\left(\frac{1+x^2}{1-x^2}\right). \label{pl4}
\end{eqnarray}
Note that the Legendre polynomial in Eq.\ (\ref{pl4}) is strictly positive because its argument is at least equal to 1.
Insertion of Eqs.\ (\ref{rg}) and \ (\ref{pl4}) 
into Eq.\ (\ref{hs1}) yields the compact formula
\begin{eqnarray}
\delta_{HS}[\hat \rho^{(M)}]&=&\frac{1}{2}+
\left(\frac{1+x}{1-x}\right)^{M+1}\frac{1}{{\cal P}_M\left(\frac{1+x^2}{1-x^2}\right)} 
\left[\frac{1}{4\langle \hat N \rangle+2}-\frac{{\langle \hat N \rangle}^M}{(\langle \hat N 
\rangle+\bar n+1)^{M+1}}\right],\nonumber \\
\label{hsf}
\end{eqnarray}
with the mean occupancy $\langle \hat N \rangle$ given by 
Eq.\ (\ref{mo}).
The situation is different for both the entropic and Bures
non-Gaussianity measures. Making use of the photon-number 
probabilities\ (\ref{pl}) and\ (\ref{sl}) in Eqs.\ (\ref{re2}) 
and\ (\ref{diag2}), we established noncompact formulae for the relative entropy and the Bures degree of non-Gaussianity. 
Each of their expressions includes a power series which has to be summed numerically. We further computed numerically these two expressions as one-parameter functions of a single variable for several values of the parameter.  

Long ago, Agarwal and Tara \cite{AT} examined the non-classicality 
of the state\ (\ref{PATS}) by writing its non-positive 
$P$ representation and Mandel's $Q$-factor. Non-Gaussianity of this state was recently evaluated in Ref.\cite{Simon} by employing the Wehrl entropy-measure and found to be equal to the non-Gaussianity 
of the number state $|M\rangle \langle M|$, being thus independent 
of the thermal mean occupancy $\bar n$. This is a consequence 
of an invariance property 
of the Wehrl entropy under a uniform phase-space scaling 
of the $Q$ function 
of the state. 

\section{Discussion and conclusions}
On physical grounds, we expect that a good measure 
of non-Gaussianity has a monotonic behaviour
with respect to the mean photon number $\langle \hat N\rangle$ and, 
in turn, to the parameters entering its expression. It is quite clear that the non-Gaussianity measures\ (\ref{hsf}),\ (\ref{re2}), 
and\ (\ref{diag2}) depend on the thermal mean occupancy $\bar n$, unlike the  Wehrl-entropy measure \cite{Simon}.
Our analytic formula, Eq.\ (\ref{hsf}), led us to accurate values 
for the Hilbert-Schmidt degree of non-Gaussianity.
 In Fig. 1 we plot the three distance-type measures as functions of the parameter $x$ for several values of the number $M$ of added photons. It is interesting that the three measures of non-Gaussianity 
$\delta_{HS}$, $\delta_{RE}$, and $\delta_F$ decrease monotonically with $x$. We did not find any extrema of these functions in contrast with Figs. 3 and 4 
in Ref.\cite{Simon}.  

\begin{figure*}[h]
\center
\includegraphics[width=5cm]{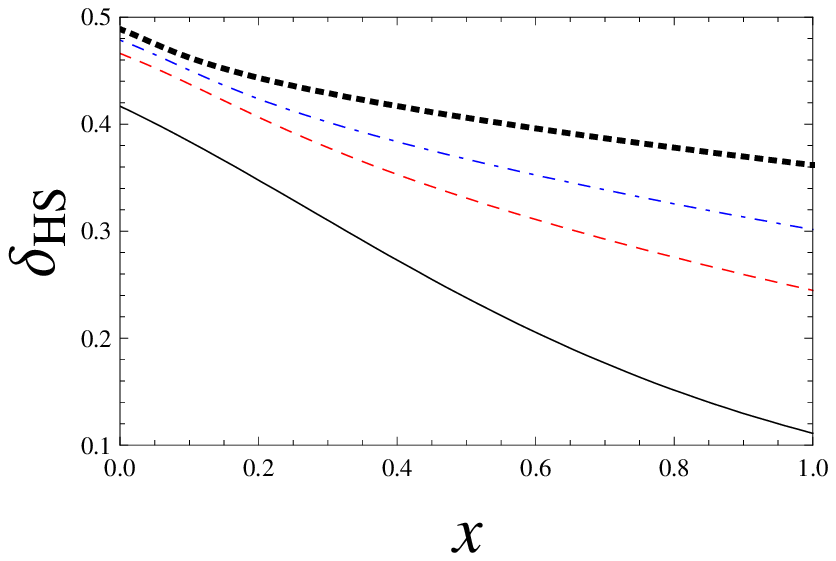}
\includegraphics[width=5cm]{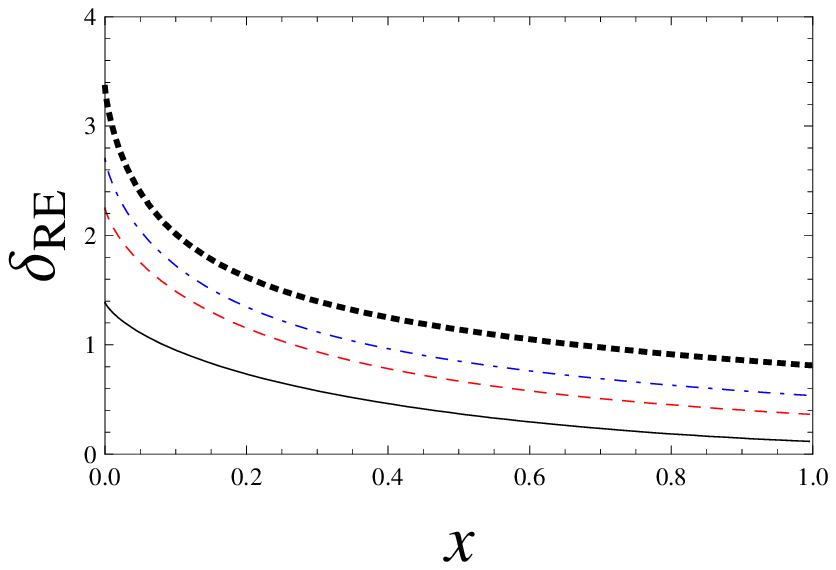}
\includegraphics[width=5cm]{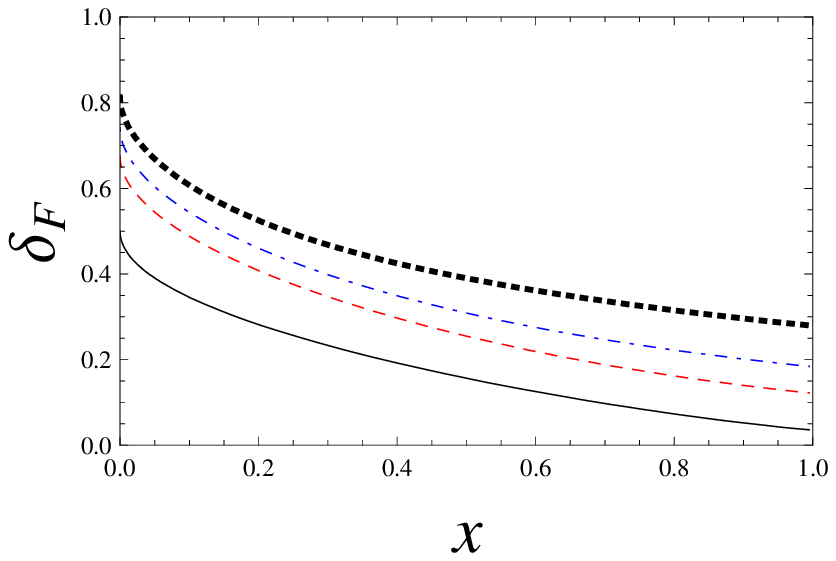}
\caption{Dependence of the distance-type non-Gaussianity on the thermal
parameter $x$ for $M$-photon-added thermal states with $M=1,3,5,10$ (from bottom to top). }
\label{fig-1}
\end{figure*}
The variation of non-Gaussianity with the number $M$ of added photons
is shown in Fig. 2 for several values of the thermal mean occupancy.
\begin{figure*}[h]
\center
\includegraphics[width=5cm]{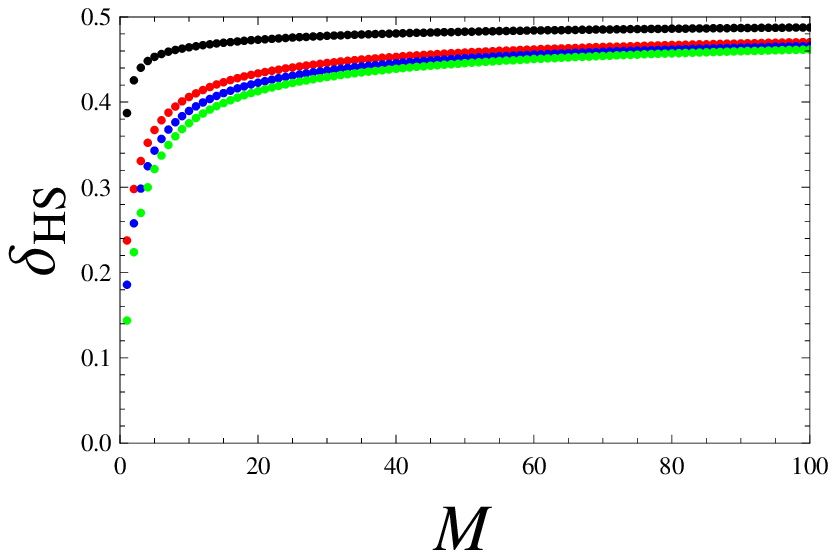}
\includegraphics[width=5cm]{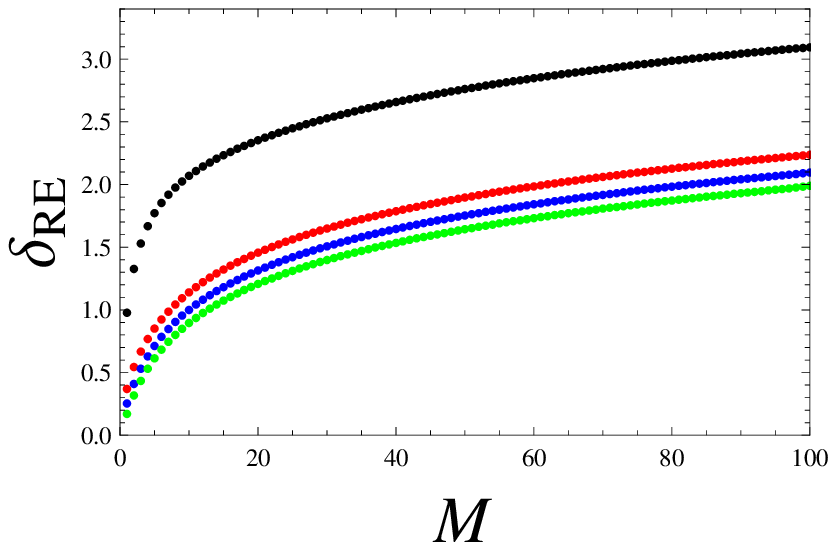}
\includegraphics[width=5cm]{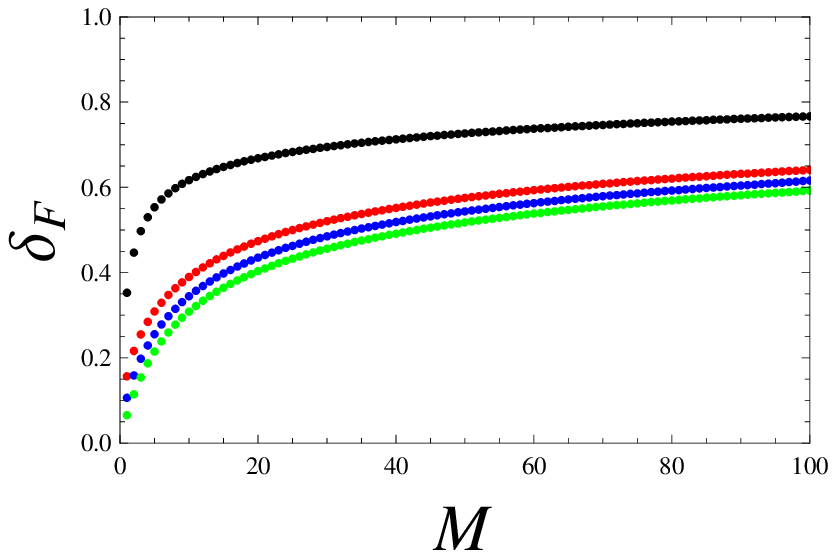}
\caption{Dependence of the distance-type non-Gaussianity on the number of added photons. All the plots start from origin. We have used $\bar n=0.1, 1, 2, 5$ (from top to bottom)}
\label{fig-2}
\end{figure*}
Besides showing a monotonic dependence on both parameters $\bar n$ 
and $M$, Figs. 1 and 2 seem to display a consistent
relation between the three non-Gaussianity measures involved.
To better outline this aspect and inspired by Ref.\cite{P33},  
we plot in Fig. 3 their mutual dependences
when the parameter $x$ varies on its domain $x \in [0,1]$ at the same
values of the number $M$ of added photons as in Fig.1. We can see 
that consistency is not present for all values of the parameters, especially in the dependence $\delta_F-\delta_{HS}$. 
\begin{figure*}[h]
\center
\includegraphics[width=5cm]{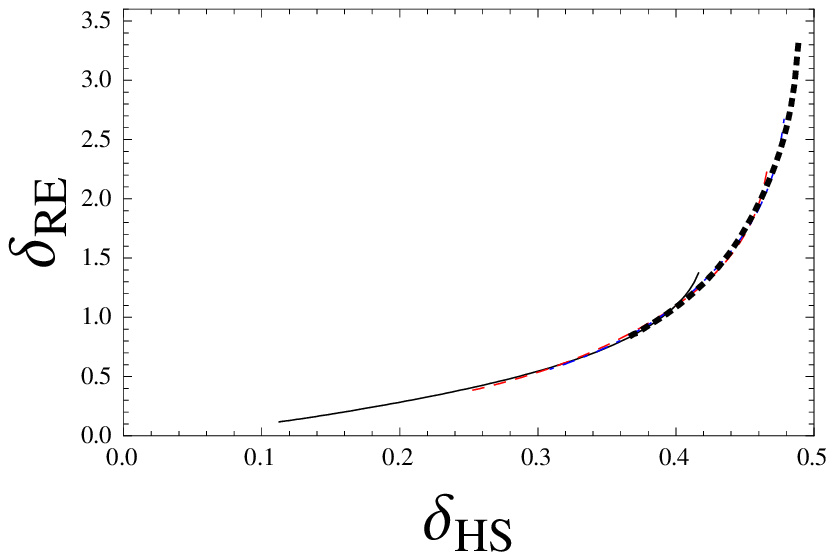}
\includegraphics[width=5cm]{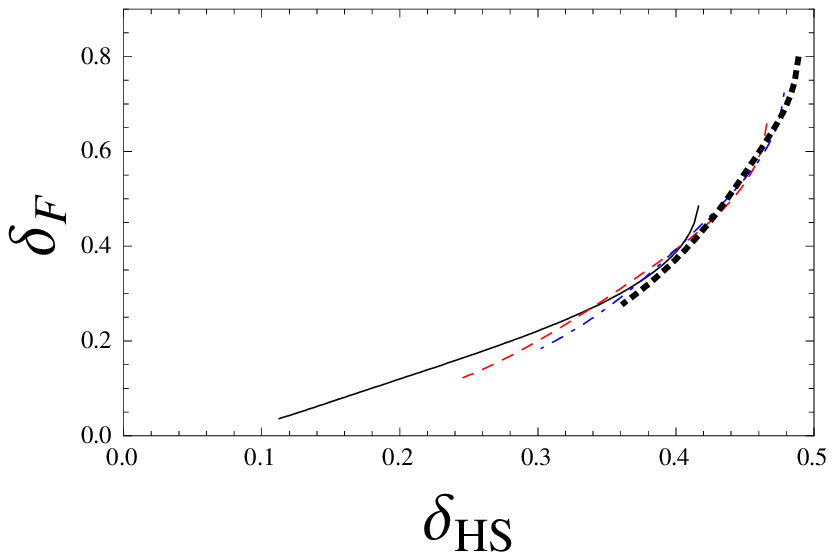}
\includegraphics[width=5cm]{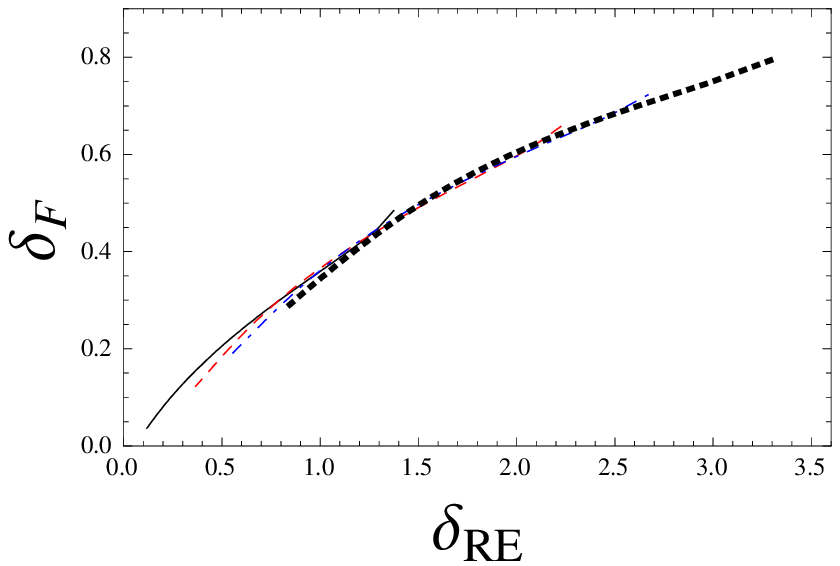}
\caption{Mutual dependence of the distance-type non-Gaussianity measures \ (\ref{hsf}),\ (\ref{re2}) and\ (\ref{diag2}) for 
$M=1,3,5,10$ (from top to bottom). We have used $\bar n\in[0,1]$. }
\label{fig-3}
\end{figure*}

To conclude, in this paper we have introduced the Bures degree 
of non-Gaussianity built with Uhlmann's fidelity between the given state and its associated Gaussian one.
We have then investigated the behaviour of three distance-type measures of non-Gaussianity for an $M$-photon-added thermal state as functions of the variables $M$ and $\bar n$. We have found 
adequate monotonic dependences of  $\delta_{HS}$, $\delta_{RE}$, 
and $\delta_F$ on both parameters $M$ and $\bar n$. This is displayed by Figs. 1 (as a function of the thermal mean occupancy) and 2 (as a function of the number of added photons). 
Although very different as geometric significance, the three measures seem to give consistent results by inducing the same ordering of non-Gaussianity. Figure 3 shows a very good consistency between 
$\delta_{RE}$ and $\delta_{HS}$ (left plot) and between $\delta_F$ and  $\delta_{RE}$ (right plot). We also notice that the plots corresponding to different numbers $M$ of added photons are 
very close for all mutual dependences.

\ack{This work was supported by the Romanian National Authority 
for Scientific Research through Grant IDEI-1012/2011 for the University of Bucharest.}

\appendix  
\section{Some useful formulae involving Gauss hypergeometric
functions}
A Gauss hypergeometric function is the sum of the corresponding hypergeometric series,
\begin{equation}
_{2}F_{1}(a, b; c ; z):=\sum\limits_{n=0}^{\infty}\frac{(a)_{n}
(b)_{n}}{(c)_{n}}\frac{z^n}{n!}\,,\qquad (|z|<1), \label{a1}
\end{equation}
where $(a)_{n}:=\Gamma(a+n)/\Gamma(a)$ is Pochhammer's symbol.
This definition is extended by analytic continuation \cite{19}.
Recall the linear transformation formula 
\begin{equation}
_{2}F_{1}(a, b; c ; z)=(1-z)^{-b}\,{_{2}F_{1}}\left(c-a, b; c ; 
\frac{z}{z-1}\right). \label{a2}
\end{equation}
The Legendre polynomial of degree $M$ can be expressed in terms 
of a Gauss hypergeometric function:
\begin{equation}{\cal P}_M(z)={_{2}F_{1}}\left(-M, M+1; 1; 
\frac{1-z}{2} \right),\qquad (M=0,1,2,3,...). \label{a3}
\end{equation}

\end{document}